\documentclass[12pt,preprint]{aastex}
\begin{document}
\newcommand{\up}[1]{\ifmmode^{\rm #1}\else$^{\rm #1}$\fi}
\newcommand{\zdot}{\makebox[0pt][l]{.}}
\newcommand{\upd}{\up{d}}
\newcommand{\uph}{\up{h}}
\newcommand{\upm}{\up{m}}
\newcommand{\ups}{\up{s}}
\newcommand{\arcd}{\ifmmode^{\circ}\else$^{\circ}$\fi}
\newcommand{\arcm}{\ifmmode{'}\else$'$\fi}
\newcommand{\arcs}{\ifmmode{''}\else$''$\fi}

\title{The Araucaria Project. Infrared Tip of the Red Giant Branch Distances to Five Dwarf 
Galaxies in the Local Group
\footnote{Based on observations obtained with the ESO NTT and VLT telescopes:
171.D-0004, 074.D-0318, 077.D-0423, 079.D-0482}
}
\author{Marek G{\'o}rski}
\affil{Warsaw University Observatory, Al. Ujazdowskie 4, 00-478, Warsaw, Poland}
\affil{Universidad de Concepci{\'o}n, Departamento de Astronomia,
Casilla 160-C, Concepci{\'o}n, Chile}
\authoremail{mgorski@astrouw.edu.pl}
\author{Grzegorz Pietrzy{\'n}ski}
\affil{Universidad de Concepci{\'o}n, Departamento de Astronomia, 
Casilla 160-C, Concepci{\'o}n, Chile}
\affil{Warsaw University Observatory, Al. Ujazdowskie 4, 00-478, Warsaw, Poland}
\authoremail{pietrzyn@astrouw.edu.pl}
\author{Wolfgang Gieren}
\affil{Universidad de Concepci{\'o}n, Departamento de Astronomia,
Casilla 160-C, Concepci{\'o}n, Chile}
\authoremail{wgieren@astro-udec.cl}

\begin{abstract}
We have obtained accurate near-infrared photometry of the Tip of the Red Giant Branches in the 
Local Group galaxies Sculptor, NGC 6822, NGC 3109, IC 1613 and WLM. We have used the derived
TRGB magnitudes together with the absolute magnitude calibration of the near-infrared TRGB magnitude
of Valenti, Ferraro and Origlia to determine the distances of these five galaxies. The statistical errors
in the distance moduli are typically 4\%. The systematic uncertainties are dominated by the knowledge
of the mean metallicities of the red giant branches, and are in the range of 5-8\%. We observe a slight
(2\%) systematic difference between the distances derived from the J and K bands, respectively, which
is within the 1 $\sigma$ errors of the distances. We compare the new distances derived in this paper
with other recent distance determinations for our target galaxies and find excellent agreement.
In particular, the near-infrared TRGB distances to the four dwarf irregular galaxies in the sample
agree to better than $5\%$ in each case with their Cepheid distances obtained from infrared photometry, indicating
that there is no appreciable systematic offset between these two fundamental techniques using
old and young stellar populations, respectively.
\end{abstract}


\section{Introduction}
The Araucaria project is a long term program designed to improve the
calibration of the cosmic distance scale by improving known, or introducing
new, stellar methods of distance measurement by extensive photometric and
spectroscopic observations of these standard candles in a number of galaxies in 
the Local Group and the nearby Sculptor Group (e.g. Gieren et al. 2005a). The
methods of distance measurement we have been using and improving in the course of the
project, both in the optical and near-infrared domains, include the Cepheid 
period-luminosity relation, the RR Lyrae period-luminosity-metallicity relation,
the mean brightness of red clump stars, and the TRGB brightness (e.g. Gieren
et al. 2005b, Szewczyk et al. 2008, Pietrzynski et al. 2009a, 2010).
We have also designed a new spectroscopic technique to measure galaxy distances
from their blue supergiant stars (Kudritzki et al. 2008), and quite recently we have
started to use late-type eclipsing binary systems in the LMC to measure a near-geometrical,
accurate distance to this anchor point of the distance scale (Pietrzynski et al. 2009b).

A crucial step for achieving accurate distance determinations is the reduction of the influence 
of weakly known error sources, like reddening and population effects, on the distance results. 
In many cases this can be achieved by using near-infrared 
observations, which has been demonstrated for red clump stars (Pietrzynski and Gieren (2002);  
Pietrzynski, Gieren and Udalski (2003)) ,  Cepheids  (Pietrzynski et al. (2006a); Gieren et al. 
(2008a, 2008b, 2009); Soszynski et al. (2006)), and for RR-Lyrae stars (Szewczyk et al. (2009) and 
Pietrzynski et al. (2008)). The TRGB method, originally introduced by Lee, Freedman and Madore (1993),
is also also an attractive tool for distance determination in the 
local universe. It marks the helium flash in old, low-mass stars arriving at the end
of their red giant phase, and in  a color-magnitude diagram (CMD) occurs at a well-defined
luminosity. It is generally accepted that the
optical I band absolute magnitude of the TRGB does not depend on age and metallicity 
under  the condition that the stars belong to an old and low metallicity population. Several studies
(e.g. Kennicutt et al. 1998; Ferrarese et al. 2000; Udalski 2000) have
confirmed this conjecture, and have shown that the TRGB brightness  is 
also insensitive to  the shape of the star formation history (SFH) for old stellar populations 
if the mean metallicity is less than [Fe/H] = -0.3 dex  (Barker et al. 2004). 
The conceptual simplicity, and the low cost in observing time together with the relatively bright absolute 
magnitude of the TRGB (about -4 mag for the I band)  
make it a useful tool for distance determination, and since the pioneering 
work of Lee et al. (1993)  the I band brightness of the TRGB  has been successfully applied to measure 
distances to most of the nearby galaxies (e.g. Ferrarese et al. 2000, Karatchentsev et al. 2003).

Important progress was achieved by Ivanov and Borissova (2002) and Valenti, 
Ferraro and Origlia (2004), who calibrated the absolute near-infrared J, H and K band magnitudes of the TRGB
in terms of the metallicity of the red giants.  
The obvious advantage of using near-infrared data is reducing the influence of the 
reddening on the distance result.  The  TRGB magnitudes in the near-infrared bands are also brighter than 
in the optical bands.  Unfortunately,  near-infrared data for nearby galaxies, 
deep enough to reliably measure the TRGB magnitudes are extremely scarce. Therefore this technique     
has so far been applied to only a very few galaxies to measure their distances  (e.g. the Magellanic Clouds 
(Cioni et al. 2000), NGC5128 (Rejkuba 2004) and the Fornax dSph galaxy (Gullieuszik et al. 2007) ). 
Recently, we have started to use the near-infrared  TRGB technique in the Araucaria project.
In a first  paper (Pietrzynski et al. 2009a) we presented near-infrared TRGB distance determinations for two Local 
Group dwarf spheroidal galaxies, Carina and Fornax. In this paper, we extend this work to five
additional Local Group galaxies for which we could collect the necessary data: the Sculptor dwarf spheroidal 
galaxy, and the irregular galaxies NGC 3109, NGC 6822, IC 1613 and WLM.

Our paper is organized as follows. The observations, reductions and calibrations of the data
are described in the following section. Next we present the distance determinations, followed by a 
discussion of the errors associated to our results and a comparison of the infrared TRGB
 distances derived in this paper with those
previously derived for these galaxies from other techniques.  Finally we present 
a summary and final remarks.

\section{Observations, Data Reduction and Calibration}
The near-infrared data presented in this paper were collected with the SOFI and ISAAC infrared 
cameras attached to the ESO NTT and VLT telescopes on La Silla and Paranal, respectively.
The field of view of SOFI (Large Field Mode)
was 4.9 x 4.9 arcmin whereas the FOV for the ISAAC images was 2.5 x 2.5  arcmin. 

During one non-photometric night 9 SOFI fields were observed in Sculptor, covering 
the main body of this galaxy. The Sculptor galaxy was already subject of earlier deep IR imaging  
performed by our group with  NTT/SOFI
(Pietrzynski et al. 2008). However, those observations were 
optimized for studying the faint RR Lyrae population in Sculptor, and the stars 
having brightnesses similar to the TRGB magnitude were saturated in
these images, or fell into the nonlinear regime of the NTT camera.
Therefore, in order to accurately measure the K and J band magnitudes of the TRGB 
in this relatively nearby galaxy we decided to follow our strategy used earlier
for the Carina and Fornax galaxies, and to observe  Sculptor again with the SOFI 
camera but now using very short individual integrations (DITs) of 1.5 s.
Such short exposure times guaranteed that the TRGB brightness was well below the non-linearity limit,
making very accurate photometry possible.
Our data were transformed onto the standard system using
about 30  common stars from the 2MASS Point Source Catalog
(Wachter et al. 2003) which were found in any given SOFI field. The scatter (rms) of the calculated
zero point offsets was always smaller than 0.01 mag.

Accurate near-infrared photometry was already obtained for Cepheid variables in  IC 1613, NGC 6822, 
WLM and NGC 3109 in the course of the Araucaria project (Pietrzynski et al. 2006, Gieren et al. 2006,
Gieren et al. 2008b, Soszynski et al. 2006). The images used for the Cepheid photometry turned out
to be very well suited for an accurate determination of the TRGB magnitudes in these galaxies. 
Indeed, in each galaxy a relatively large total area was observed ensuring 
large numbers of  RGB stars available for the photometry. The photometry is deep enough in all cases
to reach 2-3 mag below the expected magnitude of the TRGB.  More detailed information about the 
observations, and the adopted reduction and calibration procedures, can be found in the papers cited
above.

The data of  IC 1613, NGC 6822, WLM and NGC 3109 were calibrated 
onto the UKIRT system (Hawarden et al.  2001) based on extensive observations 
of standard stars. In order to transform this photometry onto the  2MASS photometric system, 
in which the calibration of the TRGB method was done, the
transformations (equation 1 and 2) derived by Carpenter (2001) were used:

${\rm {K}_{s}^{2MASS} = {K}_{UKIRT} + (0.004 \pm 0.006){(J-K)}_{UKIRT} + (0.002 \pm 0.004)}$ (1)\\

${\rm ({J}-{K}_{s})^{2MASS} = (1.069 \pm 0.011){(J-K)}_{UKIRT} + (-0.012 \pm 0.006)}$ (2)\\

\section{Distance determination}
From the photometry in the J and K bands, we have constructed  the color-magnitude diagrams (CMD) for 
our five target galaxies. Examples are given for NGC 6822 and WLM in Fig. 1. It can be appreciated
that our photometry is indeed very well suited 
to calculate accurate TRGB magnitudes. Red Giant Branch stars, selected from our CMDs were used to obtain 
the Gaussian smoothed luminosity functions, and then the
TRGB brightness was calculated with a slightly modified 
Sobel edge-detection filter (described in detail by Sakai, Madore and Freedman 1996).
In order to improve the explicitness of the filter answer, the output of the program was weighted with the
original Sobel filter set for wider sampling. This technique allows to determine the TRGB magnitude
 with great certainty and accuracy. In order to verify the reliability of this procedure,
we made comparison tests in three galaxies:  Fornax, Carina and Sculptor. 
The TRGB brightnesses in the J and K bands for these galaxies 
have been obtained with both the standard Sobel edge-detection technique, and with our modified filter, and in 
each case very good agreement between the corresponding results (within 0.04 mag) was found (see Table 2).
However, the location of the TRGB was much better pronounced and could be more 
easily and accurately identified with the modified algorithm, so we decided to adopt 
it in our project to measure the TRGB magnitudes for the target galaxies. The resulting J and K band 
brightnesses of the TRGB in our 5 galaxies are presented in Table 3. 
Exemplary luminosity functions and outputs from the modified Sobel filter are presented 
in Fig. 2 for NGC 6822 and WLM.

In order to calculate the absolute magnitudes of the TRGB in the J and K bands, we need the mean
metallicities of the red giant populations in our target galaxies.
For the Sculptor dSph galaxy, we adopt [Fe/H]=-1.83 $\pm$ 0.26  dex (Clementini et al. 2005). 
For NGC 3109, Hidalgo et al. (2008) obtained  [Fe/H]=-1.84 $\pm$ 0.2  dex, 
which is in good agreement with Minniti et al. (1999) who found [Fe/H]=-1.80 $\pm$ 0.2  dex. 
For the WLM dIrr galaxy, we adopt [Fe/H]=-1.27 $\pm$ 0.04  dex (Leaman et al. 2009).
In order to obtain the old stellar population metallicities for  IC 1613 and NGC 6822,
we used optical HST photometry from the Extragalactic Distance Database (Jacobs et al. 2009). 
Based on the colors of the RGB stars and using the  calibration of Lee, Freedman and Madore (1993), 
metallicities of [Fe/H]=-1.50 $\pm$ 0.08  dex for IC 1613, and [Fe/H]=-1.06 $\pm$ 0.09 dex for NGC 6822, 
respectively were obtained. 
Consistent metallicities were reported for NGC 6822 by Zucker and Wyder (2004) ([Fe/H]=-1.0 dex)
and IC 1613 by Lee at al. (1993) ([Fe/H] = -1.3 dex)

The Galactic foreground reddenings towards the galaxies were computed from the Schlegel, Finkbeiner and Davis (1998) 
extinction maps, and are also reported in Table 2.

To derive the absolute TRGB magnitudes for our target galaxies, we used the calibration of
Valenti, Ferraro and Origlia (2004) which was obtained from extensive near-infrared
observations of 24 Galactic globular clusters  covering a wide
metallicity range from -2.12 to -0.49 dex (equations 3 and 4), and the metallicities 
from Table 2. 

${\rm {M}_{J}^{TRGB} = -5.67 - 0.31 \times [Fe/H]}$ (3)\\

${\rm {M}_{K}^{TRGB} = -6.98 - 0.58 \times [Fe/H]}$ (4)\\

The 1 $\sigma$ scatter of the datapoints about these two calibrating relations in J and K are very similar
(0.20 mag and 0.18 mag, respectively; Valenti et al. 2004).
The absolute magnitudes calcultated from these relations, combined with our observed 
TRGB magnitudes corrected for reddening
with the values in Table 2, led to the true distance moduli of our five target galaxies reported in Table 4.

\section{Discussion}
The systematic uncertainty on each of the TRGB distance determinations in Table 4
contains contributions from the photometric zero point errors
(typically 0.02 - 0.03 mag), errors on the adopted extinctions (0.02 mag), and the uncertainty on the
adopted metallicities (typically 0.2 dex; see previous section). With the adopted metallicity dependence
of the TRGB absolute magnitude calibration (equations 3 and 4), the accuracy of the metallicity is clearly 
the dominant contribution to the total error budget, particularly in the K band for which
the metallicity sensitivity of the TRGB absolute magnitude is about twice as strong as in the J band.

The statistical error of the Sobel edge TRGB detection technique is estimated as the
Full Width at Half Maximum (FWHM) of the highest peak that marks the TRGB. 

For each target galaxy, the calculated value of the total systematic and statistic uncertainty
of the distance determination is given in Table 4. We note, however, 
that we did not take into account any contribution from possible systematic errors on the
coefficients themselves in the calibration of  Valenti, Ferraro, and Origlia (2004).

As can be seen in Table 4, the distance determinations  in the J and K bands are consistent with each other. 
The largest discrepancy occurs for IC 1613 for which the J and K band distance determinations differ by 0.12 mag 
(although this is still less than the one sigma error range). This discrepancy might be caused by some
contamination of the red giant sample close to the tip with AGB stars. 
It is worth noting that our distances obtained from K band photometry are consistently
 slightly shorter (by some 0.04 mag) than the corresponding distances
obtained from the J band photometry. While this could be a consequence of AGB star pollution affecting
the K band TRGB magnitude somewhat stronger than the J band magnitude, it could also
be related to the fact that some additional internal reddening 
is produced inside our irregular target galaxies (e.g. E(B-V) in the order of 0.1 mag) which would affect 
the J-band TRGB
magnitude somewhat stronger (e.g. make the J-band TRGB magnitude fainter) than the less affected
K-band magnitude of the tip, which is what we observe in Table 4.
 Evidence for the existence of internal reddening in our dIrr target galaxies of this order of magnitude
had come from our previous Cepheid studies, but should be mostly related to the dusty regions
in which Cepheids as young stars are embedded. It is interesting to see that the
smallest difference between the J-band and K-band distance modulus (0.02 mag) occurs for Sculptor,
the only dwarf speroidal galaxy in our sample which can be supposed to be free of intrinsic reddening.
A yet different possible explanation for a systematic difference between the distance moduli derived
from the J and K bands is coming from the uncertainties  on the slopes of the metallicity
terms in equations 3 and 4, which could likely introduce systematic deviations at the 2\% level in the 
absolute magnitudes, and hence in the distance moduli. Given the small sample of galaxies in the
present work we have for comparison, no clear-cut conclusion about the origin of a possible systematic
difference between the J- and K-band TRGB distances is possible. Studies on an enlarged sample
of galaxies might well show that there is no significant systematic difference at all.

In Table 5 we present a compilation of previous distance measurements to our target galaxies 
found in the literature. As can be seen, the near-infrared TRGB distances derived in this paper
agree very well with other recent distance measurements. The VIJK Cepheid distances
derived in the course of the Araucaria Project for the four dIrr galaxies in this study agree
in each case to better than 5\% with the infrared TRGB distances, the agreement being slightly
better with the TRGB distances coming from the J-band photometry. This very nice agreement of
a Pop. I indicator (Cepheids) with a Pop. II indicator (TRGB) is a very recomforting result
which demonstrates that genuine progress has been made in the calibration of the respective
techniques; it also demonstrates the importance of having taken the methods to the infrared
domain where reddening is not a dominant source of systematic error as it is in optical studies.

\section{Summary and Conclusions}
From near-infared photometry of stars close to the tip of the red giant branch
we have derived the distances to five Local Group galaxies using the TRGB absolute magnitude-metallicity
calibration  in the J and K bands given by Valenti, Ferraro and Origlia
(2004). Our results complement our previous distance measurements to the Carina and Fornax dwarf spheroidal galaxies 
based on this same technique (Pietrzynski et al. 2009a) and provide an opportunity 
to check on the precision of the distance determination with the TRGB method in the 
near-infrared domain. We find that the distances derived from the J and K band TRGB magnitudes agree within
one standard deviation of the results. However, the distances derived from the TRGB magnitudes in the J band
tend to have smaller systematic uncertainties because of the smaller sensitivity of the absolute J-band
TRGB magnitude on the metallicity of the red giants, as compared to the K band. On the other hand, this
advantage might be compensated by the smaller reddening sensitivity of the K-band TRGB distances when the method
is applied to galaxies containing considerable amounts of dust.

Applying the technique on dwarf irregular galaxies 
has opened the possibility to make a direct comparison with the distances derived from
Cepheids, as the fundamental Pop. I distance indicator. We find that the Cepheid and
TRGB distances derived from near-infrared photometry agree in all (4) cases to better than 5\%,
leading us to conclude that the respective calibrations of the methods are accurate at this
level, and that the infrared TRGB method of distance measurement, as a technique "cheap" in telescope
time, is an attractive tool for distance determination, and superior to the optical version
of the method because it yields distances which are basically unaffected by reddening.

\acknowledgments
We gratefully acknowledge financial support for this
work from the Chilean Center for Astrophysics FONDAP 15010003, and from
the BASAL Centro de Astrofisica y Tecnologias Afines (CATA) PFB-06/2007.
Support from the Polish grants N203 387337 and N203 509938,
and the FOCUS and TEAM subsidies of the Fundation for Polish Science (FNP) 
is also acknowledged.

\begin{figure}[p]
\vspace*{18cm}
\includegraphics{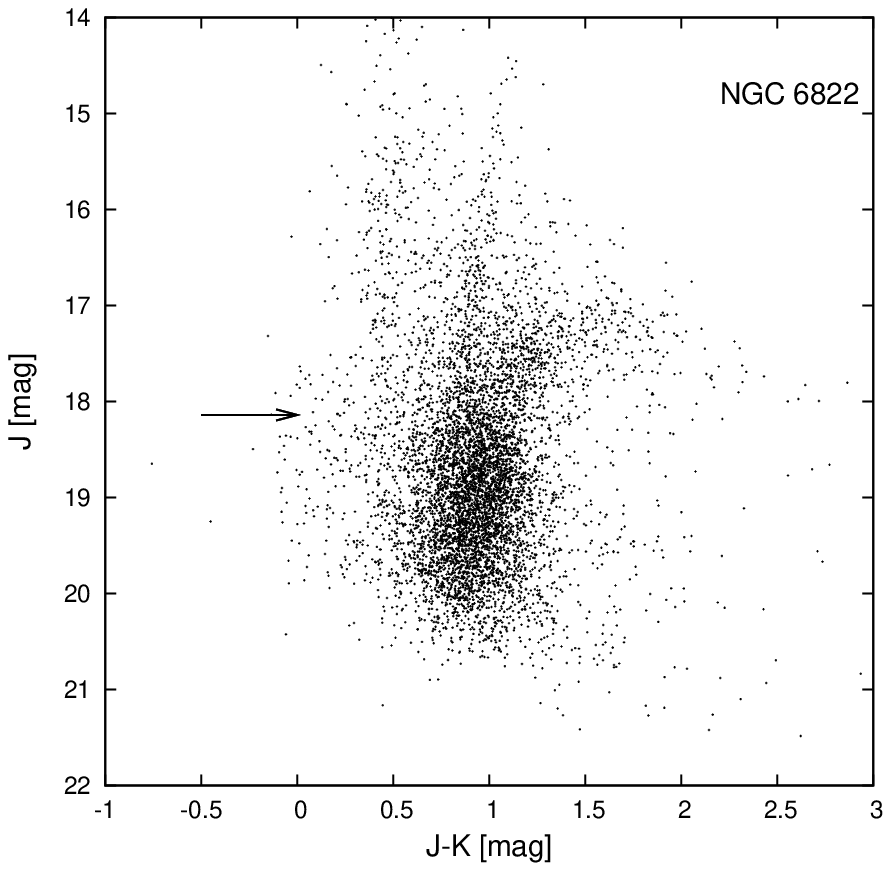}
\includegraphics{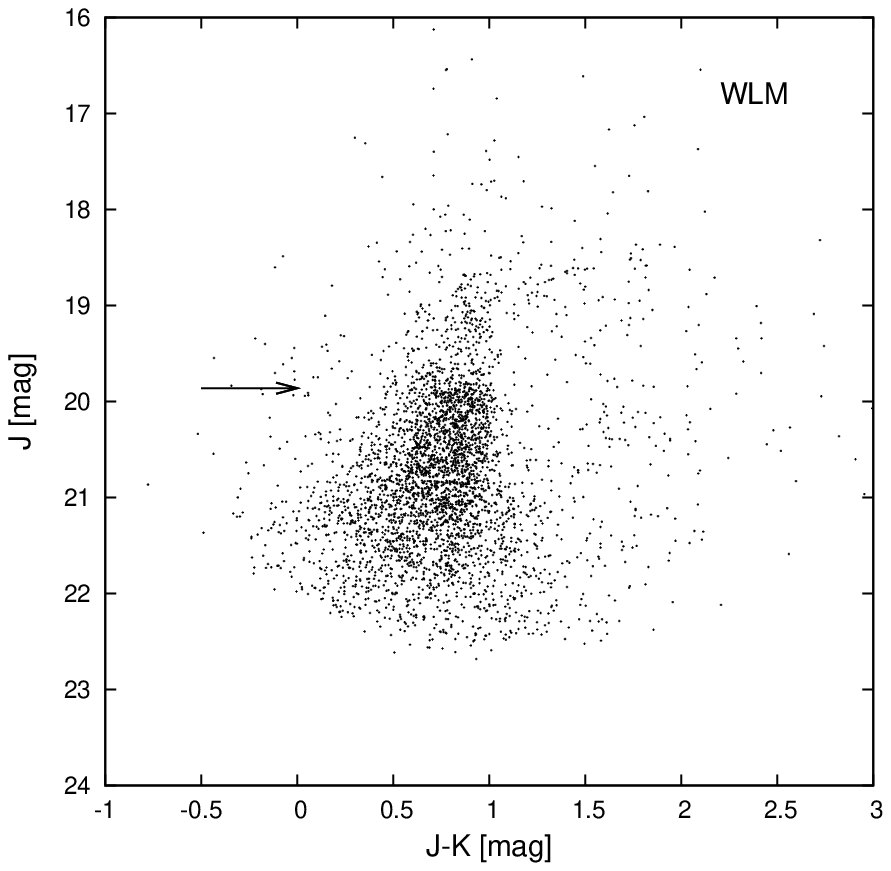}
\caption{The J, J-K color-magnitude diagrams for WLM (top) and NGC 6822 (bottom) obtained from our data. 
The arrows point at the position of the  respective TRGB magnitudes}
\end{figure}

\begin{figure}[p]
\vspace*{18cm}
\includegraphics{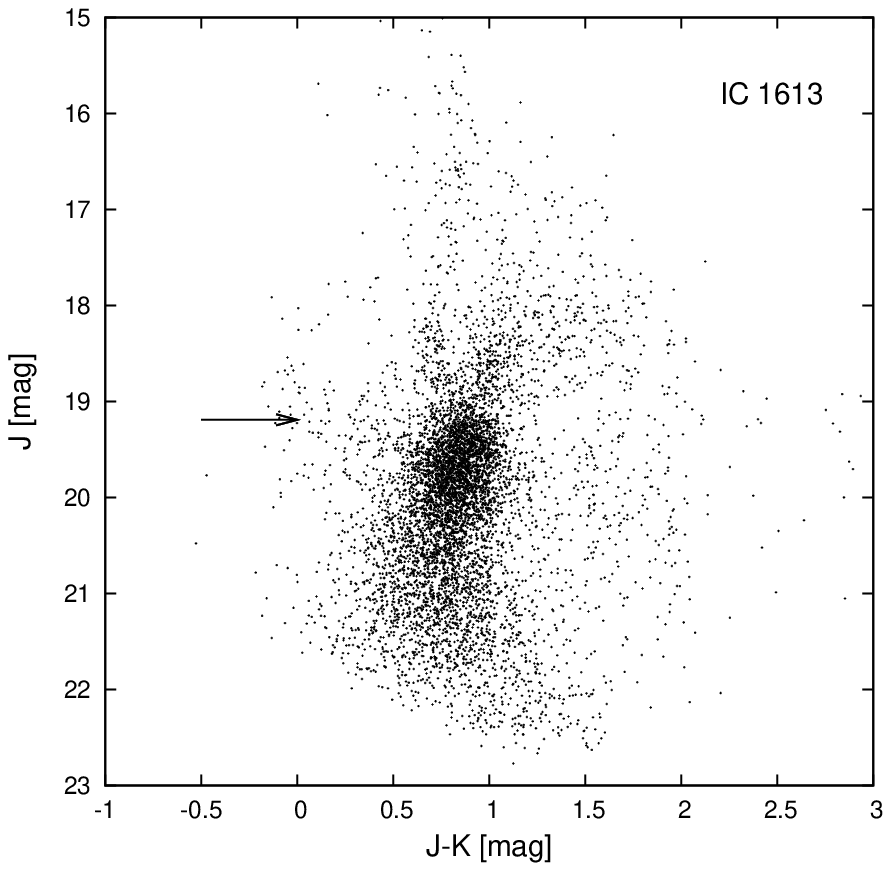}
\includegraphics{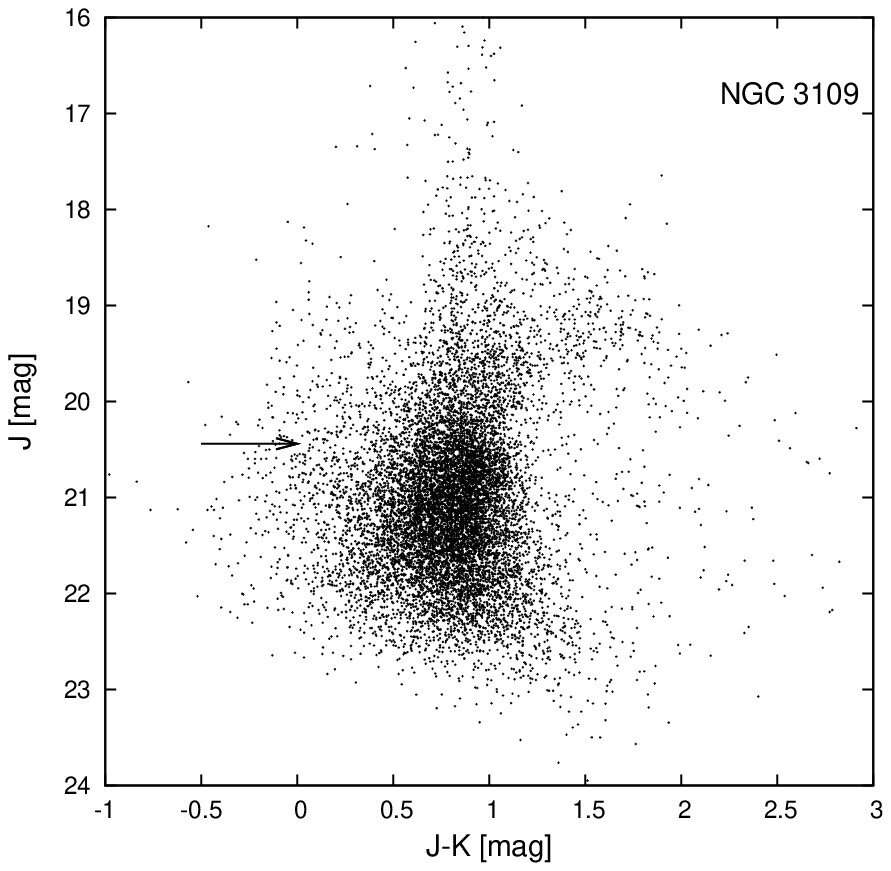}
\caption{The J, J-K color-magnitude diagrams for NGC3109 (top) and IC 1613 (bottom) obtained 
from our data. The arrows point at the position of the  respective TRGB magnitudes}
\end{figure}

\begin{figure}[p]
\vspace*{18cm}
\includegraphics{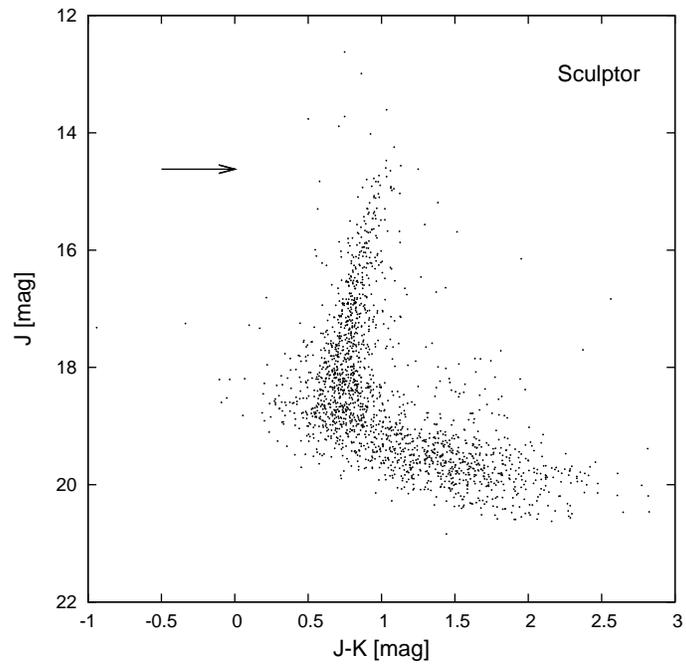}
\caption{The J, J-K color-magnitude diagrams for Sculptor, obtained from our data. 
The arrows point at the position of the  respective TRGB magnitudes}
\end{figure}

\begin{figure}[p] 
\vspace*{18cm}
\includegraphics{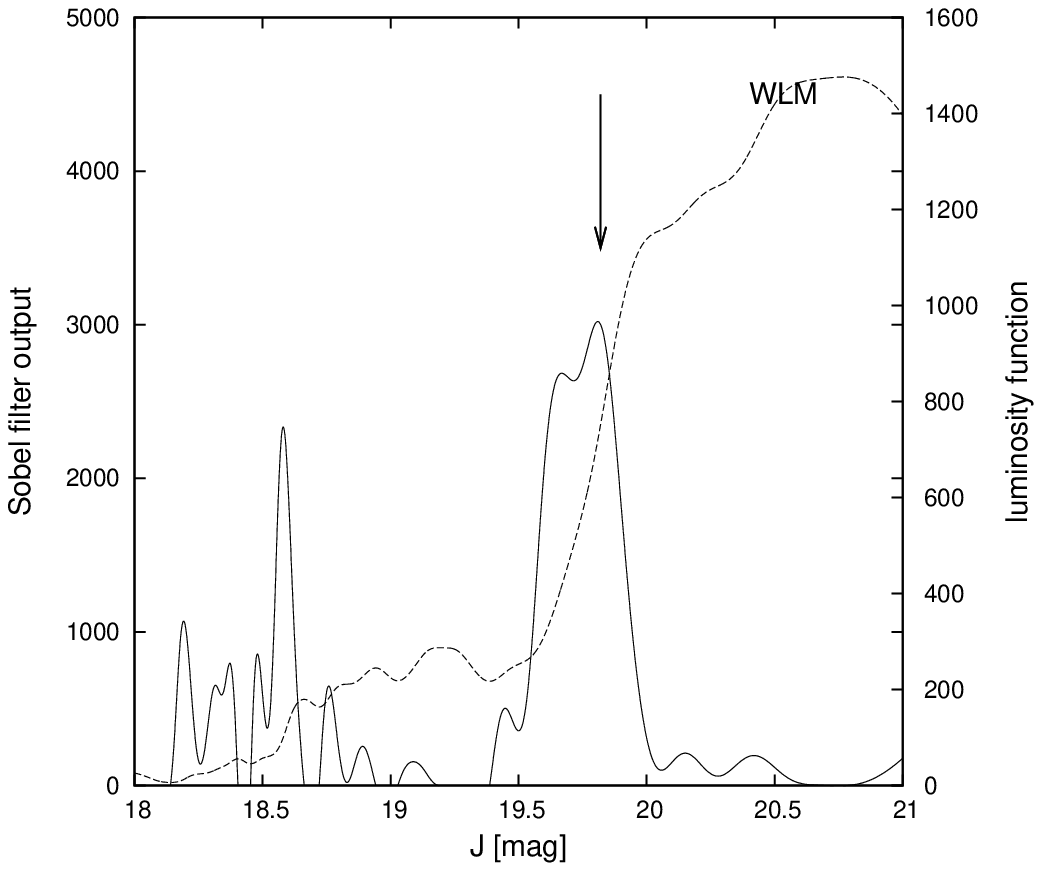}
\includegraphics{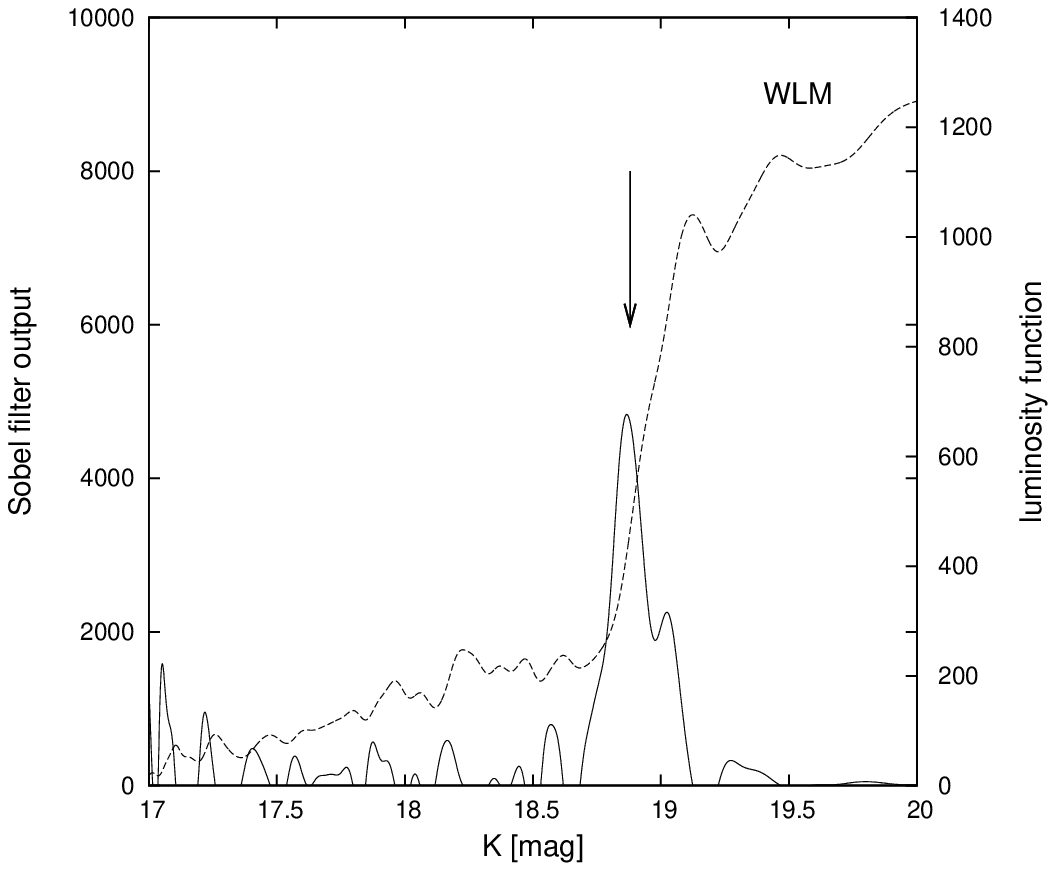} 
\includegraphics{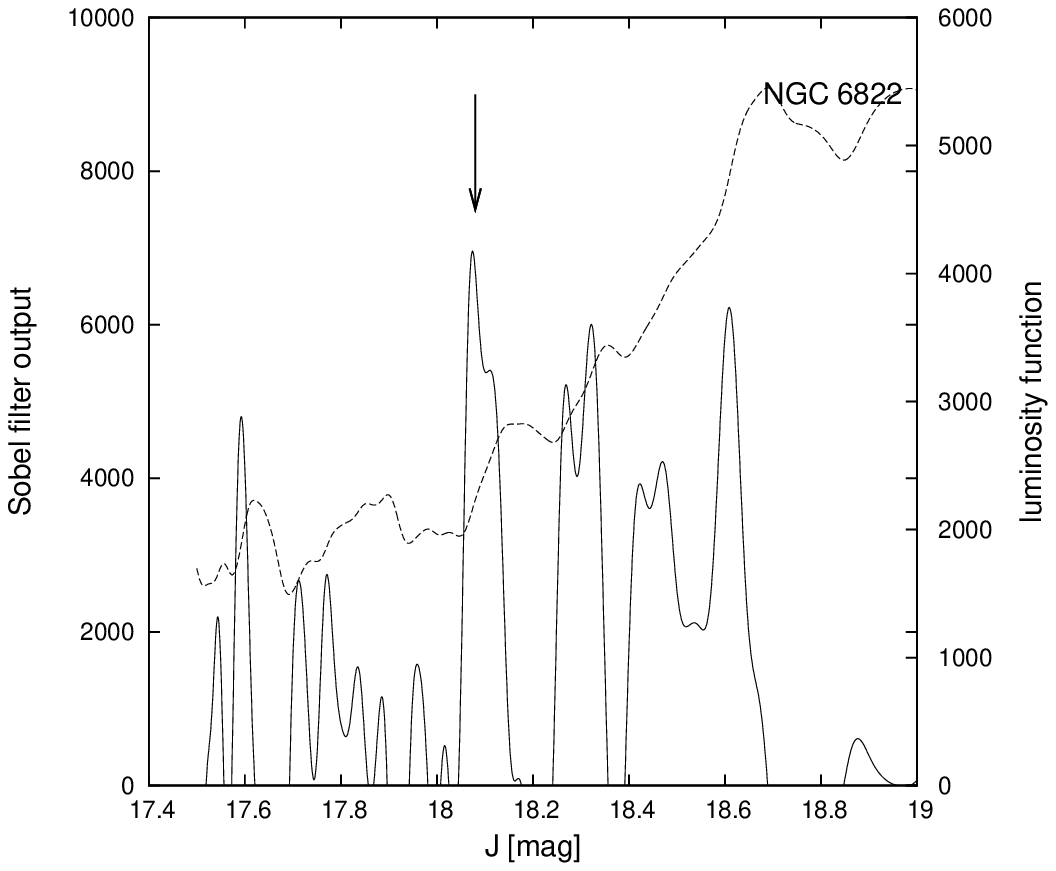}
\includegraphics{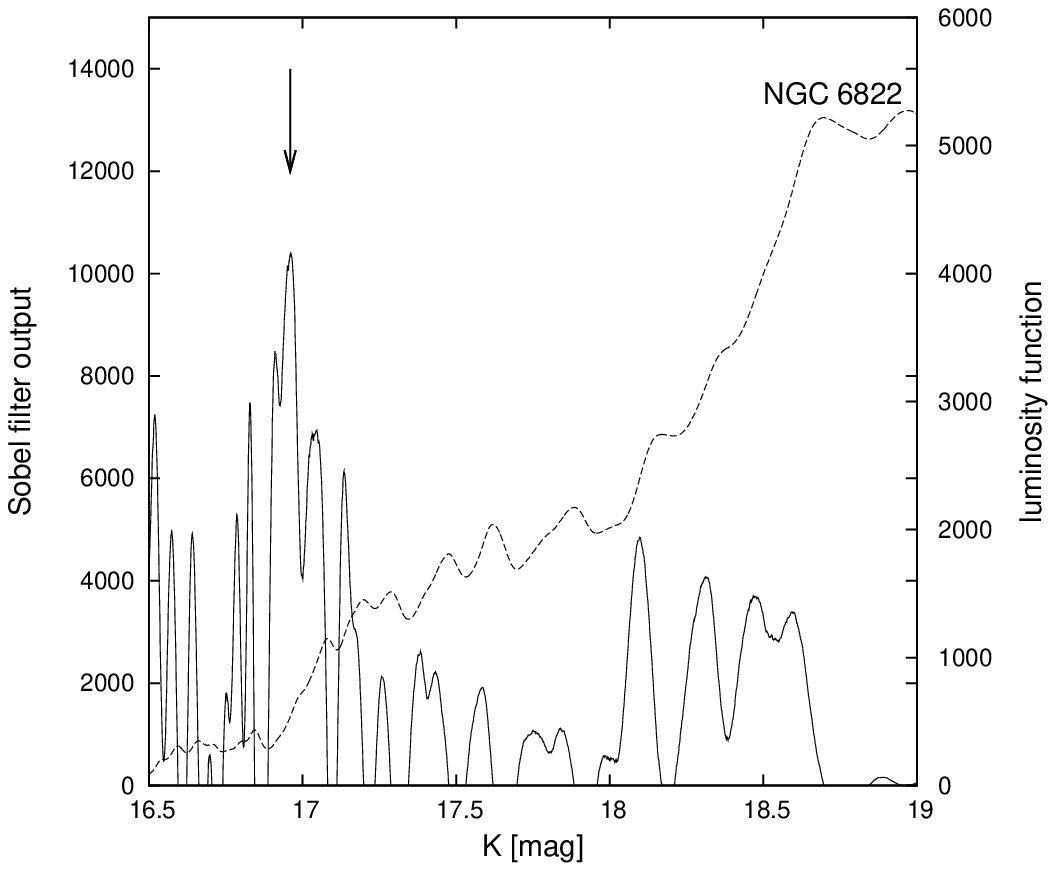}
\caption{The J and K band Gaussian-smoothed luminosity function of the red giant branch (dashed line) and the 
corresponding outputs of the edge-detection filter (solid line) in WLM  (top) and  NGC 6822  (bottom). 
The arrows indicate the detected TRGB magnitudes.}
\end{figure}  

\begin{deluxetable}{cccc}
\tablecaption{Near-infrared data}

\tablehead{\colhead{Galaxy} & \colhead{Instrument} & \colhead{Number of fields} & \colhead{Reference}\\
}
\startdata
Sculptor & SOFI & 9 & This paper \\
IC 1613  & SOFI & 8 & Pietrzynski et al. (2006a)\\
NGC 6822 & SOFI & 10 & Gieren et al. (2006)\\
NGC 3109 & ISAAC & 5 & Soszynski et al. (2006)\\
WLM      & SOFI  & 6 & Gieren et al. (2008b) \\ \hline
\enddata
\end{deluxetable}

\begin{deluxetable}{cccc}
\tablecaption{Comparison of the TRGB infrared brightnesses obtained with the
modified, and original Sobel edge-detection filters for the Carina, Fornax and Sculptor galaxies.}

\tablehead{\colhead{Galaxy} & \colhead{Band} & \colhead{Original Sobel filter} & \colhead{Modified Sobel filter}\\
\colhead{} & \colhead{} & \colhead{[mag]} & \colhead{[mag]} \\
}
\startdata
Carina   & J & 15.00 $\pm$  0.03  & 15.04 $\pm$ 0.04  \\
Carina   & K & 14.17 $\pm$  0.05  & 14.13 $\pm$ 0.09  \\
Fornax   & J & 15.51 $\pm$  0.03  & 15.58 $\pm$ 0.04  \\
Fornax   & K & 14.45 $\pm$  0.05  & 14.49 $\pm$ 0.03  \\
Sculptor & J & 14.64 $\pm$  0.06  & 14.62 $\pm$ 0.08  \\ 
Sculptor & K & 13.82 $\pm$  0.07  & 13.80 $\pm$ 0.08  \\ \hline
\enddata
\end{deluxetable}

\begin{deluxetable}{ccccc}
\tablecaption{Observed J and K band TRGB magnitudes, metallicities and reddenings}
\tablehead{\colhead{Galaxy} & \colhead{TRGB (J)} & \colhead{TRGB (K)}& \colhead{[Fe/H]} & \colhead{E(B-V)}
 \\
\colhead{} & \colhead{[mag]} & \colhead{[mag]} &
\colhead{[dex]} & \colhead{[mag]} \\
}
\startdata
IC 1613  & 19.19  $\pm$ 0.08 & 18.13  $\pm$ 0.08  & -1.50 $\pm$ 0.08 & 0.02  $\pm$ 0.02  \\

NGC 6822 & 18.14  $\pm$ 0.05 & 16.97  $\pm$ 0.09  & -1.00  $\pm$ 0.08  & 0.23  $\pm$ 0.04 \\

NGC 3109 & 20.44  $\pm$ 0.05 & 19.52  $\pm$ 0.05  & -1.84 $\pm$ 0.20 & 0.06 $\pm$ 0.02 \\

WLM      & 19.86  $\pm$ 0.10 & 18.88  $\pm$ 0.08  & -1.27 $\pm$ 0.04 & 0.04  $\pm$ 0.02 \\
   
Sculptor & 14.62  $\pm$ 0.08 & 13.80  $\pm$ 0.08 & -1.83 $\pm$ 0.26 & 0.02   $\pm$ 0.02\\ \hline
\enddata
\end{deluxetable}

%
%
%

\begin{deluxetable}{ccccccc}
\tablecaption{Distance moduli obtained for the target galaxies from their infrared TRGB magnitudes}
\tablehead{ \colhead{Galaxy} & \colhead{ $(m-M)_{0}^{J}$} & \colhead{$\sigma$ (stat.)} & \colhead{$\sigma$ (syst.)} & \colhead{$(m-M)_{0}^{K}$} & \colhead{$\sigma$ (stat.)}  & \colhead{$\sigma$ (syst.)} \\
\colhead{} & \colhead{[mag]} & \colhead{[mag]} &  \colhead{[mag]} &  \colhead{[mag]} & \colhead{[mag]} &  \colhead{[mag]}\\
}
\startdata
IC 1613   & 24.36 & 0.08 & 0.09 & 24.24 & 0.08 & 0.10 \\
NGC 6822  & 23.31 & 0.05 & 0.10 & 23.26 & 0.07 & 0.10 \\
NGC 3109  & 25.49 & 0.05 & 0.09 & 25.42 & 0.05 & 0.13 \\
WLM       & 25.14 & 0.09 & 0.12 & 25.12 & 0.08 & 0.15 \\
Sculptor  & 19.72 & 0.08 & 0.11 & 19.70 & 0.08 & 0.17 \\ \hline
\enddata
\end{deluxetable}

\begin{deluxetable}{cccccc}
\tablecaption{Distance determinations for target galaxies 
from different techniques}
\tablehead{\colhead{Galaxy} & \colhead{Method} & \colhead{Band}&
\colhead{Distance Modulus} & Error & Reference \\
 \\
\colhead{} & \colhead{} & \colhead{} &
\colhead{[mag]} & \colhead{[mag]} & \\
}
\startdata
IC1613  & Cepheid & Mid-IR & 24.27 & 0.02 & Freedman et al. (2009) \\
IC1613 & Cepheid & JK & 24.291 & 0.035 & Pietrzynski et al. (2006a) \\
IC 1613 & Cepheid & V,I & 24.20 & 0.07 & Udalski et al. (2001) \\
IC1613 & RR Lyrae & V,I & 24.31 & 0.06 & Dolphin et al. (2001) \\
IC1613 & TRGB & I & 24.38 & 0.05 & Jacobs et al. (2009) \\
IC1613 & TRGB & J & 24.12 & 0.25  & Jung et al. (2009)\\
IC1613 & TRGB & H & 24.20 & 0.44 & Jung et al. (2009)\\
IC1613 & TRGB & K & 24.00 & 0.52 & Jung et al. (2009)\\
\hline
NGC6822 & Cepheid & JK & 23.312 & 0.021 & Gieren et al. (2006) \\
NGC6822 & Cepheid & Mid-IR & 23.49 & 0.03 & Madore et al. (2009) \\
NGC6822 & Cepheid & V,I & 23.34 & 0.06 & Pietrzynski et al. (2004) \\
NGC6822 & RR Lyrae & V & 23.36 & 0.18 & Clementini et al. (2003)\\
NGC6822 & TRGB    & I   & 23.34  & 0.12 & Cioni and Habing (2005) \\
NGC6822 & TRGB & J & 23.35 & 0.26 & Sohn et al. (2008) \\
NGC6822 & TRGB & H & 23.20 & 0.42 & Sohn et al. (2008) \\
NGC6822 & TRGB & K & 23.27 & 0.50 & Sohn et al. (2008) \\
\hline
NGC3109 & Cepheid & JK & 25.571 & 0.024 & Soszynski et al. (2006) \\
NGC3109 & Cepheid & V & 25.72 & 0.03 & Pietrzynski et al. (2006b) \\
NGC3109 & Cepheid & I & 25.66 & 0.03 & Pietrzynski et al. (2006b) \\
NGC3109 & TRGB & I & 25.45 & 0.15 & Lee (1993) \\
\hline
WLM & Cepheid & V,I & 25.144 & 0.07 & Pietrzynski et al. (2007) \\
WLM & Cepheid & JK & 24.925 & 0.042 & Gieren et al. (2008b) \\
WLM & FGLR & Spectr. & 24.99 & 0.1 & Urbaneja et al. (2008) \\
WLM & HB   &  V      &  24.95 & 0.13 & Rejkuba et al. (2000) \\
WLM & TRGB & I & 24.81 & - & Lee, Freedman and Madore (1993)\\
WLM & TRGB & I & 24.85 & 0.08 & McConnachie et al. (2005) \\
WLM & Cepheid & I & 24.92 & - & Lee, Freedman and Madore (1993)\\
\hline
Sculptor & RR Lyrae & J,K & 19.67 & 0.12 & Pietrzynski et. al. (2008) \\
Sculptor & TRGB & optical & 19.64 & 0.08 & Rizzi (2002) \\
Sculptor & HB & V & 19.66 & 0.15 & Rizzi (2002) \\
Sculptor & RR Lyrae& V  & 19.71 & 0.18 & Kaluzny et al. (1995)\\
\hline
\enddata
\end{deluxetable}


\begin{references}

\reference{} Barker, M.K., Sarajedini, A., Harris, J., 2004, \apj, 606, 869

\reference{} Carpenter, J. M., 2001, \aj, 121, 2851

\reference{} Cioni, M.R., van der Marel, R.P., Loup, C. and Habing, H.J., 2000, \aap, 359, 601

\reference{} Cioni, M.R.,  and Habing, H.J., 2005, A\&A, 429, 837

\reference{} Clementini, G, Held, E.V., Baldacci, L., Rizzi, L., 2003, \apj, 588, L85

\reference{} Clementini, G., et al. 2005, \mnras, 363, 734

\reference{} Dolphin, A.E., Saha, A., Skillman, E.D., Tolstoy, E., Cole, A.A., Dohm-Palmer, R.C., 
Gallagher, J.S., Mateo, M., Hoessel, J.G., 2001, \apj, 550, 554

\reference{} Ferrarese, L., et al., 2000, \apjs, 128, 431

\reference{} Freedman, W.L., Rigby, J., Madore, B.F., Persson, S.E., Sturch, L., Mager, V., 2009, \apj, 695, 996

\reference{} Gieren, W., Pietrzy{\'n}ski, G., Bresolin, F., et al., 2005a, 
Messenger, 121, 23

\reference{} Gieren, W., Pietrzy{\'n}ski, G., Soszy{\'n}ski, I., Bresolin, F., Kudritzki, R.P.,
Minniti, D., Storm, J., 2005b, \apj, 628, 695

\reference{} Gieren, W., Pietrzy{\'n}ski, G., Nalewajko, K., Soszy{\'n}ski, I., 
Bresolin, F., Kudritzki, R.P., Minniti, D., and Romanowsky, A., 2006, \apj, 647, 1056

\reference{} Gieren, W., Pietrzy{\'n}ski, G., Soszy{\'n}ski, I., Bresolin, F.,
Kudritzki, R.P., Storm, J. and Minniti, D., 2008a, \apj, 672, 266

\reference{} Gieren, W., Pietrzy{\'n}ski, G., Szewczyk, O., Soszy{\'n}ski, I.,
Bresolin, F., Kudritzki, R.P., Urbaneja, M.A., Storm, J. and Minniti, D., 2008b,
\apj, 683, 611

\reference{} Gieren, W., Pietrzy{\'n}ski, G., Soszy{\'n}ski, I., Szewczyk, O., 
Bresolin, F., Kudritzki, Urbaneja, M.A., Storm and Garcia-Varela, A., 2009, \apj, 700, 1141 


\reference{} Hawarden, T.G., Leggett, S.K., Letawsky, M.B., Ballantyne, D.R., Casali, M.M., 2001, \mnras, 325, 563 

\reference{} Hidalgo, S.L., Aparicio, A., Gallart, C., 2008, \aj, 136, 2332

\reference{} Ivanov, V.D., Borissova, J., 2002, \aap, 390, 937

\reference{} Jacobs, B.A., Rizzi, L., Tully, R.B., Shaya, E.J., Makarov, D.I., and Makarova, L., 2009, \aj, 138, 332

\reference{} Jung, M.Y., Chun, S.-H., Chang, C.-R., Han, M., Lim, D., Han, W., Sohn, Y.-J, 2009, JASS, 26, 421

\reference{} Kaluzny, J., Kubiak, M., Szymanski, M., Krzeminski, W., and Mateo, M., 1995, \aapr, 112, 40

\reference{} Karatchentsev, I., et al., 2003, \aap, 404, 93

\reference{} Kennicutt, R.C.Jr., et al., 1998, \apj, 498, 181



\reference{} Kudritzki, R.P., Urbaneja, M.A., Bresolin, F., Przybilla, N., Gieren, W., 
Pietrzy{\'n}ski, G., 2008, \apj, 681, 269

\reference{} Leaman, R., Cole, A.A., Venn, K.A., Tolstoy, E., Irwin, M.J., Szeifert, T., Skillman, E.D., McConnachie, A.W., \apj, 699, 1

\reference{} Lee, M.G., Freedman, W.L., Madore, B.F., 1993, \apj, 417, 553

\reference{} Lee, M.G., 1993, \apj, 408, 409


\reference{} Madore, B.F., Rigby, J., Freedman, W.L., Persson, S.E., Sturch, L., Mager, V., 2009, \apj, 693, 946

\reference{} McConnachie, A.W., Irwin, M.J., Ferguson, A.M.N., Ibata, R.A., Lewis, G.F., Tanvir, N., \mnras, 
356, 979

\reference{} Minniti, D., Zijlstra, A.A., Alonso, M.V., 1999, \aj, 117, 881

\reference{} Minniti, D., Zijlstra, A.A., 1997, \aj, 114, 147


\reference{} Pietrzy{\'n}ski, G., and Gieren, W., 2002, \aj, 124, 2633

\reference{} Pietrzy{\'n}ski, G., Gieren, W., and Udalski, A., 2003, \aj, 125, 2494

\reference{} Pietrzy{\'n}ski, G., Gieren, W., Udalski, A., Bresolin, F.,
Kudritzki, R.-P., Soszy{\'n}ski, I., Szymo{\'n}ski, M., Kubiak, M., 2004, \aj, 128, 2815

\reference{} Pietrzy{\'n}ski, G., Gieren, W., Soszy{\'n}ski, I., Bresolin, F., 
Kudritzki, R.-P.,Dall'Ora, M., Storm, J., and Bono, G., 2006a, \apj, 642, 216

\reference{} Pietrzy{\'n}ski, G., Gieren, Udalski, A., W., Soszy{\'n}ski, I., Bresolin, F.,
Kudritzki, R.-P., Mennickent, R., Kubiak, M., Szymo{\'n}ski, M., Hidalgo, S., 2006b, \apj, 648, 366

\reference{} Pietrzy{\'n}ski, G., Gieren, W., Udalski, A., Soszy{\'n}ski, I., Bresolin, F.,
Kudritzki, R.P., Garcia, A., Minniti, D., Mennickent, R., Szewczyk, O., Szymanski, M., Kubiak, M.,
Wyrzykowski, L., 2007, \aj, 134, 594

\reference{} Pietrzy{\'n}ski, G., Gieren, W., Szewczyk, O., Walker, A.,
Rizzi, L., Bresolin, F., Kudritzki, R.-P., Nalewajko, K., Storm, J.,Dall'Ora, M., 
Ivanov, V., 2008, \aj, 135, 1993

\reference{} Pietrzy{\'n}ski, G., G{\'o}rski, M., Gieren, W., Ivanov, V. D., Bresolin, F., 
and Kudritzki, R.-P., 2009a, \aj, 138, 459

\reference{} Pietrzy{\'n}ski, G., Thompson, I.B., Graczyk, D., Gieren, W., Udalski, A., Szewczyk, O.,
Minniti, D., Kolaczkowski, Z., Bresolin, F., Kudritzki, R.P., 2009b, \apj, 697, 862

\reference{} Pietrzy{\'n}ski, G., Gorski, M., Gieren, W., Laney, D., Udalski, A., 2010, \aj, 140, 1038

\reference{} Rejkuba, M., 2004, A\&A, 413, 903

\reference{} Rejkuba, M., Minniti, D., Gregg, M.D., Zijlstra, A.A., Alonso, M.V., Goudfrooij, P, 2000, \aj, 120, 801 

Gullieuszik, M., 2007, \mnras, 380, 1255

\reference{} Rizzi, L., 2002, PhD Thesis, Padova University

\reference{} Sakai, S., Madore, B., Freedman, W.L., 1996, \apj, 461, 713


\reference{} Schlegel, D.J., Finkbeiner, D.P., and Davis, M., 1998,
\apj, 500, 525

\reference{} Sohn, Y.-J., Kang, A., Han, W., Park, J.-H., Kim, H.-I., Shin, J.-W., Chun, S.-H., 2008, JASS, 25, 249

\reference{} Soszy{\'n}ski, I., Gieren, W., Pietrzy{\'n}ski, G.,
Bresolin, F., Kudritzki,R.P., and Storm, J., 2006, \apj, 648, 375

\reference{} Szewczyk, O., Pietrzy{\'n}ski, G., Gieren, W., Storm, J., Walker, A.R., Rizzi, L., Kinemuchi, K.,
Bresolin, F., Kudritzki, R.P., Dall'Ora, M., 2008, \aj, 136, 272

\reference{} Szewczyk, O., Pietrzynski, G., Gieren, Ciechanowska, A., Bresolin, F., Kudritzki,
R.-P., 2009, \aj, 138, 1661


\reference{} Udalski, A., 2000, Acta Astron., 50, 279

\reference{} Udalski, A., Wyrzykowski, L., Pietrzynski, G., Szewczyk, O., Szymanski, M., Kubiak, M., 
Soszynski, I., Zebrun, K.,  2001, Acta Astron., 51, 221

\reference{} Urbaneja, M.,A., Kudritzki, R.P., Bresolin, F., Przybilla, N., Gieren, W. 
and Pietrzy{\'n}ski, G., 2008, ApJ, 684, 118

\reference{} Valenti, E., Ferraro, F.R., Origlia, L., 2004, \mnras, 354, 815

\reference{} Wachter, S., Hoard, D., Hansen, K.H., Wilcox, R.E., Taylor, H.M., Finkelstein, S.L., 2003, \apj, 586, 1356

\reference{} Zucker, D.B., Wyder, T.K. 2004, Origin and Evolution of the Elements, from the Carnegie Observatories Centennial Symposia. 
Carnegie Observatories Astrophysics Series
\end{references}
\end{document}